\documentclass[epj]{svjour}
\usepackage{amssymb}
\usepackage{amsmath}
\usepackage[dvips,final]{graphicx}
\usepackage{epsfig}
\usepackage{bm}
\usepackage{graphics}
\begin{document}
\title{Soft end-point and mass corrections to the $\eta ^{\prime}g^{\ast
}g^{\ast }$ vertex function}
\author{S. S. Agaev \inst{1,}\thanks{E-mail address: agaev$\_$ shahin@yahoo.com (Corresponding author)}
 \and M. A. Gomshi Nobary\inst{2,}\thanks{E-mail address: mnobary@razi.ac.ir}% etc
% \thanks is optional - remove next line if not needed
%\thanks{\emph{E-mail address: agaev_shahin@yahoo.com}
}%
                    % Do not remove
%

\institute{Institute for Physical Problems, Baku State University,
Z. Khalilov st.\ 23, Az-1148 Baku, Azerbaijan \and Department of
Physics, Faculty of Science, Razi University, Kermanshah, Iran}
\date{Received: 25 June 2007 / Revised version: 5 November 2007/}
% The correct dates will be entered by Springer
%
\abstract{ Power-suppressed corrections arising from end-point
integration regions to the space-like vertex function of the
massive $\eta ^{\prime }$-meson virtual gluon transition $\eta
^{\prime }-g^{\ast }g^{\ast }$ are computed. Calculations are
performed within the standard hard-scattering approach (HSA) and
the running coupling method supplemented by the infrared
renormalon calculus. Contributions to the vertex function from the
quark and gluon contents of the $\eta ^{\prime }$-meson are taken
into account and the Borel resummed expressions for $F_{\eta
^{\prime }g^{\ast }g^{\ast }}(Q^{2},\omega ,\eta )$, as well as
for $F_{\eta ^{\prime }gg^{\ast }}(Q^{2},\omega =\pm 1,\eta )$ and
$F_{\eta ^{\prime }g^{\ast }g^{\ast }}(Q^{2},\omega =0,\eta )$ are
obtained. It is demonstrated that the power-suppressed corrections
$\sim (\Lambda ^{2}/Q^{2})^{n}$, in the
explored range of the total gluon virtuality $1~\mathrm{\leq Q^{2}\leq 25\,{%
GeV}^{2}}$, considerably enhance the vertex function relative to
the results found in the framework of the standard HSA with a
fixed\ coupling. Modifications generated by the $\eta ^{\prime
}$-meson mass effects are discussed.
 % end of PACS codes
} %end of abstract
\authorrunning {S. S. Agaev and M. A. Gomshi Nobary }
\titlerunning {Corrections to the $\eta^{\prime}g^{*}g^{*}$ vertex function}
\maketitle
%
%\PACS{PACS-key}{12.38.Bx, 14.40.Aq, 11.10.Hi}
\authorrunning {S. S. Agaev and M. A. Gomshi Nobary}
\titlerunning {Soft end-point and mass corrections}
\maketitle

\section{Introduction}

\label{sec:int}

Recently interest in theoretical investigations of the gluonic
structure of the $\eta $ and $\eta ^{\prime }$-mesons has risen
due to the high precision CLEO results on the electromagnetic
$\eta \gamma ,~\eta ^{\prime }\gamma $ transition form factors
(FFs) \cite{cleo1}, as well as because of the observed large
branching ratios for the exclusive $B\rightarrow K+\eta ^{\prime
}$ and semi-inclusive $B\rightarrow \eta ^{\prime }+ X_{s}$ decays
\cite{cleo2,cleo201}.

The data on FF of the $\eta ^{\prime }\gamma $ transition were
mainly used for extracting constraints on the quark component of
the $\eta ^{\prime }$-meson distribution amplitude (DA)
\cite{FK98,FK9801,Ag01}. In these investigations various
theoretical schemes and methods were employed. An important
conclusion drawn from these studies is that the quark component of
the $\eta ^{\prime }$-meson DA should be close to its asymptotic
form and that the admixture of the first non-asymptotic term
should be within the range of $B_{2}^{q}(1 \, \rm{GeV}^2)\simeq
0.05-0.15$, $B_{2}^{q}$ being the first Gegenbauer coefficient.

An effect of the gluon component of the $\eta ^{\prime }$-meson DA
on the $\eta ^{\prime }\gamma $ transition was analyzed in
\cite{Ag03,Pas02}, where relevant constraints on the input
parameters $B_{2}^{q}$ and $B_{2}^{g}$ were extracted: it was
shown that their allowed values are strongly correlated. Useful
bounds on the Gegenbauer coefficients $B_{2}^{q}$ and $B_{2}^{g}$
were obtained also from investigation of the semi-inclusive decay
$\Upsilon(1 S) \to \eta^{\prime}+X$ \cite{par}.

 The two-gluon valence Fock component of the $\eta ^{\prime
}$-meson can directly contribute to the $\eta ^{\prime }\gamma $
transition FF only at the next-to-leading order due to quark box
diagrams and also affect the leading order result through
evolution of the quark component of the $\eta ^{\prime }$-meson
DA. Hence, an effect of the $\eta ^{\prime }$-meson gluon
component on the $\eta ^{\prime }\gamma $ transition is mild.
Contrary, the contribution of the gluon content of the $\eta
^{\prime }$-meson to the two-body non-leptonic exclusive and
semi-inclusive decay ratios of the $B$-meson may be sizeable.
Indeed, in order to explain the observed large branching ratio
$\mathrm{Br}(B\rightarrow \eta ^{\prime }+X_{s})$, in \cite{ATW} a
mechanism that employs the two-gluon content of the $\eta ^{\prime
}$-meson was suggested. In accordance with this approach the
dominant fraction of the $B\rightarrow \eta ^{\prime }+X_{s}$
decay rate appears as the result of the transition $g^{\ast
}\rightarrow g\eta ^{\prime }$ of a virtual gluon from the
standard model penguin diagram $b\rightarrow sg^{\ast}$. In
\cite{ATW} the $g^{\ast }g\eta ^{\prime }$ vertex function (VF)
was approximated by the constant $H(q^{2},0,m_{\eta ^{\prime
}}^{2})\simeq H(0,0,m_{\eta ^{\prime }}^{2})\simeq
1.8~\mathrm{GeV}^{-1}$, the latter being extracted from the
analysis of the $J/\psi \rightarrow \eta ^{\prime }\gamma $ decay.
Further investigations demonstrated that the effects of the QCD
running coupling $\alpha _{\mathrm{s}}(q^{2})$ \cite{HT}, as well
as the momentum dependence of the form factor $H(q^{2},0,m_{\eta
^{\prime }}^{2})$, properly taken into account, considerably
reduce the contribution to $\mathrm{Br}(B\rightarrow \eta ^{\prime
}+X_{s})$ of the mechanism under consideration \cite{KP}. To
eliminate the discrepancy between theoretical predictions and the
experimental data in \cite{AKS} a gluon fusion mechanism was
proposed. In accordance with the latter, the $\eta ^{\prime }$
-meson is produced by the fusion of a gluon from the QCD penguin
diagram $b\rightarrow sg^{\ast }$ with another one emitted by the
light quark inside the $B$-meson. In this mechanism, the vertex
function $F_{\eta ^{\prime }g^{\ast }g^{\ast
}}(q{_{1}}^{2},q_{2}^{2},m_{\eta ^{\prime }}^{2})$ appears owing
to the $g^{\ast }g^{\ast }\rightarrow \eta ^{\prime }$ transition.
Similar ideas form a basis for the computation of the branching
ratios of various two-body non-leptonic exclusive decay modes and
transition FFs of the $B$ meson
\cite{Ali98,Ali9801,Li,Li01,Li02,Eeg,Eeg01}.

Hence, the $\eta ^{\prime }$-meson virtual (on-shell) gluon
transition VF, $F_{\eta ^{\prime }g^{\ast }g^{\ast
}}(q_{1}^{2},q_{2}^{2},m_{\eta ^{\prime }}^{2})$, is the central
ingredient of the relevant analysis performed within perturbative
QCD (pQCD) and it deserves further investigations. This VF was
computed in numerous works \cite{Pas02,Muta,Ag02,Ali03}. The
space-like massless $\eta ^{\prime }g^{\ast }g^{\ast }$ vertex
within the standard hard-scattering approach\\ (HSA) was
considered in \cite{Pas02}, where correct analysis of the
normalization of the gluon component of the $\eta ^{\prime
}$-meson DA and that of the gluon projector onto a pseudoscalar
meson state was performed. Power-suppressed corrections, arising
from the end-point integration regions $x\rightarrow 0,1$, to the
massless space-like $\eta ^{\prime }g^{\ast }g^{\ast }$ VF were
found in \cite{Ag02}. In this work the standard HSA and the
running coupling (RC) method together with the infrared (IR)
renormalon calculus were applied. The\ $\eta ^{\prime }$-meson
mass corrections to the $\eta ^{\prime }g^{\ast }g^{\ast }$ space-
and time-like vertex in the standard HSA were calculated in \cite
{Ali03}.

In the present work, we extend the results obtained in \cite{Ag02}
by taking into account the $\eta ^{\prime }$-meson mass effects,
which may be considerable. The RC method \cite{Ag95,Ag9501}
enables us to estimate power corrections coming from the end-point
$x\rightarrow 0,1$ regions in the integrals determining the
amplitude of the $\eta ^{\prime }g^{\ast }g^{\ast } $ transition.
Indeed, in the framework of the standard HSA \cite{BL,BL01,BL02},
in order to calculate the amplitude of the process, one has to
perform integrations over the longitudinal momentum fractions of
the constituents of the meson. If one chooses the renormalization
scale $\mu _{R}^{2}$ in the hard-scattering amplitude $T_{H}$ of
the corresponding partonic subprocess in such a way as to minimize
higher-order corrections and allows the QCD coupling constant
$\alpha _{\mathrm{s}}(\mu _{R}^{2})$ to run, then one encounters
divergences arising from the end-point $x\rightarrow 0,1$ regions.
The reason is that the scale $\mu _{R}^{2}$ , as a rule, is equal
to the momentum squared of the hard virtual partons carrying the
strong interactions in the subprocess' Feynman diagrams and
depends, in general, on $x$ (or $\overline{x}\equiv 1-x$). Within
the RC method this problem is resolved by applying the
renormalization group equation and the IR renormalon calculus (for
a review see, \cite{Ben,Ben01}). It turns out that such treatment
allows us to evaluate power corrections to the physical quantity
under consideration \cite{Ag01,Ag03,Ag02,Ag98,Ag04,AGP}.

This paper is structured as follows. In Sect.\ \ref{sec:eta} we present the
necessary information on the quark-gluon structure of the $\eta ^{\prime }$%
-meson, its DAs and the hard-scattering amplitudes of the relevant
subprocesses. Sect.\ \ref{sec:quark} is devoted to calculation of
the quark component of the $\eta ^{\prime }g^{\ast }g^{\ast }$
vertex function. The contribution to the VF of the gluon content
of the $\eta^{\prime}$-meson is computed in Sect.\
\ref{sec:gluon}. Section\ \ref{sec:num} contains our numerical
results and their analysis. In Sect.\ \ref{sec:conc} we make our
concluding remarks.

\section{ Quark and gluon content of the $\protect\eta ^{\prime}$-meson and
the $\protect\eta ^{\prime }g^{\ast }g^{\ast }$ vertex}

\label{sec:eta}

\setcounter{equation}0

The Fock state decomposition of the pseudoscalar $P=\eta ,\,\eta
^{\prime }$-mesons can be written in the following form

\begin{equation*}
\left\vert P\right\rangle =\left\vert P_{a}\right\rangle +\left\vert
P_{b}\right\rangle +\left\vert P_{c}\right\rangle +\left\vert
P_{g}\right\rangle ,
\end{equation*}%
where $\left\vert P_{a}\right\rangle $ and $\left\vert P_{b}\right\rangle $
denote the $P$-meson light quarks, and $\left\vert P_{c}\right\rangle $ and $%
\left\vert P_{g}\right\rangle $ its charm and gluon components, respectively.

The light-quark content of the $P$-meson can be described either in the $%
SU_{f}(3)$ octet-singlet or in the quark-flavor basis. In this paper we
choose to work in the quark-flavor basis
\begin{equation}
\left\vert \eta _{q}\right\rangle =\frac{\Psi _{q}}{\sqrt{2}}\left\vert u%
\overline{u}+d\overline{d}\right\rangle ,\,\,\,\,\,\left\vert \eta
_{s}\right\rangle =\Psi {_{s}}\left\vert \overline{s}\right\rangle .
\label{eq:1.1}
\end{equation}%
Here $\Psi _{i}$ denote wave functions of the corresponding parton states.

We neglect the charm component of the $\eta ^{\prime }$-meson,
because in accordance with existing estimations
\cite{Pol,Pol01,Pol02}, it is too small to affect considerably the
$B$-meson exclusive decays.

The pure light-quark sector of the $\eta-\eta ^{\prime }$ system
without charm and gluon admixtures can be treated as
superpositions of the basic states (\ref{eq:1.1}),
\begin{equation*}
\left\vert \eta \right\rangle =\cos \phi _{p}\left\vert \eta
_{q}\right\rangle -\sin \phi _{p}\left\vert \eta _{s}\right\rangle ,
\end{equation*}%
\begin{equation}
\left\vert \eta ^{\prime }\right\rangle =\sin \phi _{p}\left\vert \eta
_{q}\right\rangle +\cos \phi _{p}\left\vert \eta _{s}\right\rangle .
\label{eq:1.2}
\end{equation}%
One of the advantages of the quark-flavor basis is that in this basis the
decay constants $f_{P}^{q(s)}$ follow with great accuracy the pattern of the
state mixing \cite{F00}
\begin{equation*}
f_{\eta }^{q}=f_{q}\cos \phi _{p},\,\,\,f_{\eta }^{s}=-f_{s}\sin \phi _{p},
\end{equation*}%
\begin{equation}
f_{\eta ^{\prime }}^{q}=f_{q}\sin \phi _{p},\,\,\,f_{\eta ^{\prime
}}^{s}=f_{s}\cos \phi _{p},  \label{eq:1.3}
\end{equation}%
where the decay constants $f_{q}$ and $f_{s}$, and the mixing angle $\phi
_{p}$ have the values
\begin{equation*}
f_{q}=(1.07\pm 0.02)f_{\pi },\,\,\ f_{s}=(1.34\pm 0.06)f_{\pi },
\end{equation*}
\begin{equation}
\phi _{p}=39.3^{\circ }\pm 1.0^{\circ },  \label{eq:1.4}
\end{equation}%
with $f_{\pi }=0.131\,\mathrm{GeV}$ being the pion weak decay constant.

The singlet part of the $\eta ^{\prime }$-meson DA, which is only
relevant to our present investigations, depends on both the quark
$\phi ^{q}(x,\mu
^{2})$ and gluon $\phi ^{g}(x,\mu ^{2})$ components of the $\eta ^{\prime }$%
-meson DA. These functions satisfy the symmetry and antisymmetry conditions
under the exchange $x\leftrightarrow \overline{x}$,
\begin{equation}
\phi ^{q}(x,\mu ^{2})=\phi ^{q}(\overline{x},\mu ^{2}),\,\,\,\,\,\phi
^{g}(x,\mu ^{2})=-\phi ^{g}(\overline{x},\mu ^{2}),  \label{eq:1.5}
\end{equation}%
and they are given by the expressions
\begin{equation*}
\phi ^{q}(x,\mu _{F}^{2})=6Cx\overline{x}\left\{ 1+\sum_{n=2,4..}^{\infty }%
\left[ B_{n}^{q}\left( \frac{\alpha _{\mathrm{s}}(\mu _{0}^{2})}{\alpha _{%
\mathrm{s}}(\mu _{F}^{2})}\right) ^{\frac{\gamma _{+}^{n}}{\beta _{0}}%
}\right. \right.
\end{equation*}
\begin{equation}
\left. \left. +\rho _{n}^{g}B_{n}^{g}\left( \frac{\alpha _{\mathrm{s}}(\mu
_{0}^{2})}{\alpha _{\mathrm{s}}(\mu _{F}^{2})}\right) ^{\frac{\gamma _{-}^{n}%
}{\beta _{0}}}\right] C_{n}^{3/2}(x-\overline{x})\right\}  \label{eq:1.6}
\end{equation}
and

\begin{equation*}
\phi ^{g}(x,\mu _{F}^{2})=Cx\overline{x}\sum_{n=2,4..}^{\infty }\left[ \rho
_{n}^{q}B_{n}^{q}\left( \frac{\alpha _{\mathrm{s}}(\mu _{0}^{2})}{\alpha _{%
\mathrm{s}}(\mu _{F}^{2})}\right) ^{\frac{\gamma _{+}^{n}}{\beta _{0}}%
}\right.
\end{equation*}
\begin{equation}
\left. +B_{n}^{g}\left( \frac{\alpha _{\mathrm{s}}(\mu _{0}^{2})}{\alpha _{%
\mathrm{s}}(\mu _{F}^{2})}\right) ^{\frac{\gamma _{-}^{n}}{\beta _{0}}}%
\right] C_{n-1}^{5/2}(x-\overline{x}),  \label{eq:1.7}
\end{equation}
where the constant $C$ is defined as
\begin{equation*}
C=\sqrt{2}f_{q}\sin \phi _{p}+f_{s}\cos \phi _{p}.
\end{equation*}

In (\ref{eq:1.6}) and (\ref{eq:1.7}), $C_{n}^{3/2}(z)$ and $
C_{n-1}^{5/2}(z)$ are Gegenbauer polynomials, $\mu _{F}^{2}$ \ and
$\mu _{0}^{2}$ are the factorization and normalization scales,
respectively. The values of the input parameters $B_{n}^{q}$ and
$B_{n}^{g}$ have to be fixed at the normalization scale $\mu
_{0}^{2}=1\ \mathrm{GeV}^{2}$: they
determine the shape of the DAs. In the above expressions, $\alpha _{\mathrm{s%
}}(\mu ^{2})$ is the QCD coupling constant in the two-loop approximation
given by
\begin{equation}
\alpha _{\mathrm{s}}(\mu ^{2})=\frac{4\pi }{\beta _{0}\ln (\mu ^{2}/\Lambda
^{2})}\left[ 1-\frac{2\beta _{1}}{\beta _{0}^{2}}\frac{\ln \ln (\mu
^{2}/\Lambda ^{2})}{\ln (\mu ^{2}/\Lambda ^{2})}\right] ,  \label{eq:1.9}
\end{equation}%
with $\beta _{0}$ and $\beta _{1}$ being the one- and two-loop coefficients
of the QCD beta function
\begin{equation}
\beta _{0}=11-\frac{2}{3}n_{f},\,\,\,\beta _{1}=51-\frac{19}{3}n_{f}.
\label{eq:1.10}
\end{equation}%
Here, $\Lambda $ is the QCD scale parameter and $n_{f}$ is number of active
quark flavors.

In this work, we shall use the $\eta ^{\prime }$-meson DA that
contains only the first non-asymptotic terms. Stated differently,
we suppose that in (\ref{eq:1.6}) and (\ref{eq:1.7})
$B_{2}^{q}\neq 0,\,\,B_{2}^{g}\neq 0$ and $B_{n}^{q}=B_{n}^{g}=0$
for all $n\geq 4$. Taking into account the expressions for the
required Gegenbauer polynomials as well as the values of the
relevant parameters, we can recast the $\eta ^{\prime }$-meson
quark and gluon DAs into the following simple forms \cite{Ag02}
\begin{equation*}
\phi ^{q}(x,\mu _{F}^{2})=6Cx\overline{x}\left[ 1+A(\mu _{F}^{2})-5A(\mu
_{F}^{2})x\overline{x}\right] ,
\end{equation*}
\begin{equation}
\phi ^{g}(x,\mu _{F}^{2})=Cx\overline{x}(x-\overline{x})B(\mu _{F}^{2}).
\label{eq:1.11}
\end{equation}
For $n_{f}=4$ the functions $A(\mu _{F}^{2})$ and $B(\mu _{F}^{2})$ are
defined by
\begin{equation*}
A(\mu _{F}^{2})=6B_{2}^{q}\left( \frac{\alpha _{\mathrm{s}}(\mu _{F}^{2})}{%
\alpha _{\mathrm{s}}(\mu _{0}^{2})}\right) ^{\frac{48}{75}}-\frac{B_{2}^{g}}{%
17}\left( \frac{\alpha _{\mathrm{s}}(\mu _{F}^{2})}{\alpha _{\mathrm{s}}(\mu
_{0}^{2})}\right) ^{\frac{107}{75}},
\end{equation*}%
\begin{equation}
B(\mu _{F}^{2})=19B_{2}^{q}\left( \frac{\alpha _{\mathrm{s}}(\mu _{F}^{2})}{%
\alpha _{\mathrm{s}}(\mu _{0}^{2})}\right) ^{\frac{48}{75}}+5B_{2}^{g}\left(
\frac{\alpha _{\mathrm{s}}(\mu _{F}^{2})}{\alpha _{\mathrm{s}}(\mu _{0}^{2})}%
\right) ^{\frac{107}{75}}.  \label{eq:1.12}
\end{equation}%
The $\eta ^{\prime }$-meson quark and gluon DAs for $n_{f}=3$ can
be found in \cite{Ag02}.

The $\eta ^{\prime }$-meson virtual gluon transition vertex function $%
F_{\eta ^{\prime }g^{\ast }g^{\ast }}(Q^{2},\omega ,\eta )$ is the sum of
the quark and gluon components defined in terms of the invariant amplitudes
for the process
\begin{equation*}
\eta ^{\prime }(P)\rightarrow g^{\ast }(q_{1})+g^{\ast }(q_{2}),
\end{equation*}%
in the following way:
\begin{equation}
M^{q(g)}=-iF_{\eta ^{\prime }g^{\ast }g^{\ast }}^{q(g)}(Q^{2},\omega ,\eta
)\delta _{ab}\epsilon ^{\mu \nu \rho \sigma }\epsilon _{\mu }^{a\ast
}\epsilon _{\nu }^{b\ast }q_{1\rho }q_{2\sigma }.  \label{eq:1.13}
\end{equation}%
Here, $\epsilon _{\mu }^{a},\,\,\epsilon _{\nu }^{b}$ and $q_{1},\;q_{2}$
are the polarization vectors and four-momenta of the two gluons,
respectively. Because we study only the space-like VF, $q_{1}^{2}$ and $%
q_{2}^{2}$ obey the constraints $Q_{1}^{2}=-q_{1}^{2}\geq 0$ and $%
\,\,Q_{2}^{2}=-q_{2}^{2}\geq 0$. The VF, $F_{\eta ^{\prime }g^{\ast }g^{\ast
}}(Q^{2},\omega ,\eta )$, depends on the total gluon virtuality $Q^{2}$, the
asymmetry parameter $\omega $, and $\eta ^{\prime }$ -meson scaled mass $%
\eta $
\begin{equation}
Q^{2}=Q_{1}^{2}+Q_{2}^{2},\,\ \,\,\omega =\frac{Q_{1}^{2}-Q_{2}^{2}}{Q^{2}}%
,\ \ \eta =\frac{m_{\eta ^{\prime }}^{2}}{Q^{2}}.  \label{eq:1.14}
\end{equation}%
The parameter $\omega $ varies in the region $-1\leq \omega \leq 1$. The
values $\omega =\pm 1$ corresponds to the $\eta ^{\prime }$-meson on-shell
gluon transition and $\omega =0$ to the situation when the gluons have equal
virtualities.

In accordance with the factorization theorems of pQCD, at high momentum
transfer the components of the VF $F_{\eta ^{\prime }g^{\ast }g^{\ast
}}^{q(g)}(Q^{2},\omega ,\eta )$ can be calculated by means of the formulas
\begin{equation*}
F_{\eta ^{\prime }g^{\ast }g^{\ast }}^{q}(Q^{2},\omega ,\eta )=\int_{0}^{1}
\left[ T_{1}^{q}(x,Q^{2},\omega ,\eta ,\mu _{F}^{2})\right.
\end{equation*}
\begin{equation}
\left. +T_{2}^{q}(x,Q^{2},\omega ,\eta ,\mu _{F}^{2})\right] \phi ^{q}(x,\mu
_{F}^{2})dx,  \label{eq:1.15}
\end{equation}
and
\begin{equation*}
F_{\eta ^{\prime }g^{\ast }g^{\ast }}^{g}(Q^{2},\omega ,\eta )=\int_{0}^{1}
\left[ T_{1}^{g}(x,Q^{2},\omega ,\eta ,\mu _{F}^{2})\right.
\end{equation*}
\begin{equation}
\left. -T_{2}^{g}(x,Q^{2},\omega ,\eta ,\mu _{F}^{2})\right] \phi ^{g}(x,\mu
_{F}^{2})dx.  \label{eq:1.16}
\end{equation}

\begin{figure}[tbp]
%%Fig 1
\epsfxsize=9 cm \epsfysize=6 cm \centerline{\epsffile{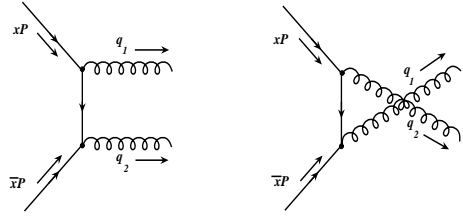}}
\vspace{-1.5cm}
\caption{Leading-order Feynman diagrams
contributing to the hard-scattering subprocess
$q+\overline{q}\rightarrow g^{\ast }+g^{\ast }$.}
\label{fig:quark}
\end{figure}

\begin{figure}[tbp]
%%Fig 2
\epsfxsize=9 cm \epsfysize=6 cm \centerline{\epsffile{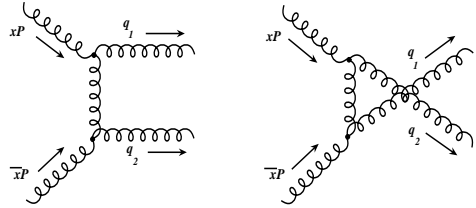}}
\vspace{-1.5cm}
\caption{Feynman diagrams contributing at leading order to the subprocess $%
g+g\rightarrow g^{\ast }+g^{\ast }$. }
\label{fig:gluon}
\end{figure}

The sum
\begin{equation*}
T_{H}^{q}(x,Q^{2},\omega ,\eta ,\mu _{F}^{2})=T_{1}^{q}(x,Q^{2},\omega ,\eta
,\mu _{F}^{2})
\end{equation*}%
\begin{equation*}
+T_{2}^{q}(x,Q^{2},\omega ,\eta ,\mu _{F}^{2})
\end{equation*}%
and the difference
\begin{equation*}
T_{H}^{g}(x,Q^{2},\omega ,\eta ,\mu _{F}^{2})=T_{1}^{g}(x,Q^{2},\omega ,\eta
,\mu _{F}^{2})
\end{equation*}%
\begin{equation*}
-T_{2}^{g}(x,Q^{2},\omega ,\eta ,\mu _{F}^{2})
\end{equation*}%
are the hard-scattering amplitudes of the subprocesses $q+\overline{q}%
\rightarrow g^{\ast }+g^{\ast }$ and $g+g\rightarrow g^{\ast
}+g^{\ast }$, respectively. The Feynman diagrams contributing at
the leading order to these subprocesses are depicted in Figs.\
\ref{fig:quark} and \ref{fig:gluon}.
%%%%%%%%%%%%%%%%%%%%%%%%%%%%%%%%%%%%%%%%%%%%%%%%%%%%%%%%%%%%%%%%%%%%%
%                            F I G U R E  1                         %
%                          \label{fig:quark}                        %
%%%%%%%%%%%%%%%%%%%%%%%%%%%%%%%%%%%%%%%%%%%%%%%%%%%%%%%%%%%%%%%%%%%%%
%%%%%%%%%%%%%%%%%%%%%%%%%%%%%%%%%%%%%%%%%%%%%%%%%%%%%%%%%%%%%%%%%%%%%
%%%%%%%%%%%%%%%%%%%%%%%%%%%%%%%%%%%%%%%%%%%%%%%%%%%%%%%%%%%%%%%%%%%%%
%                            F I G U R E  2                         %
%                          \label{fig:gluon}                        %
%%%%%%%%%%%%%%%%%%%%%%%%%%%%%%%%%%%%%%%%%%%%%%%%%%%%%%%%%%%%%%%%%%%%%
%%%%%%%%%%%%%%%%%%%%%%%%%%%%%%%%%%%%%%%%%%%%%%%%%%%%%%%%%%%%%%%%%%%%%

At the leading order of pQCD, the hard-scattering amplitudes do
not depend on the factorization scale $\mu _{F}^{2}$, but depend
implicitly on the renormalization scale $\mu _{R}^{2}$ through
$\alpha _{\mathrm{s}}(\mu _{R}^{2})$. As the scales $\mu _{R}^{2}$
and $\mu _{F}^{2}$ are independent of each other and can be chosen
separately, we adopt in this work the standard\ choice for the
factorization scale $\mu _{F}^{2}=Q^{2}$, and we omit in what
follows the dependence of the hard-scattering amplitudes on
$\mu_{F}^{2}$. Thus, we have
\begin{equation*}
T_{1}^{q}(x,Q^{2},\omega ,\eta ,\mu _{R}^{2})=-\frac{2\pi }{3Q^{2}}\frac{%
\alpha _{\mathrm{s}}(\mu _{R}^{2})}{\omega \lambda }
\end{equation*}
\begin{equation}
\times \frac{\omega (1+\lambda )-\eta (x-\overline{x})}{x(1+\omega )+%
\overline{x}(1-\omega )+2x\overline{x}\eta },  \label{eq:1.17}
\end{equation}
and
\begin{equation*}
T_{1}^{g}(x,Q^{2},\omega ,\eta ,\mu _{R}^{2})=\frac{\pi \alpha _{\mathrm{s}%
}(\mu _{R}^{2})}{Q^{2}n_{f}}
\end{equation*}
\begin{equation}
\times \frac{x(1+\omega )+\overline{x}(1-\omega )+2(1+x\overline{x})\eta }{%
\omega \lambda \left[ \overline{x}(1+\omega )+x(1-\omega )+2x\overline{x}%
\eta \right] },  \label{eq:1.18}
\end{equation}
where $\lambda =(1+2\eta /\omega ^{2}+\eta ^{2}/\omega
^{2})^{1/2}$ \cite{Ali03}. The remaining two functions can be
obtained from (\ref{eq:1.17}) and (\ref{eq:1.18}) by means of the
replacement $x\leftrightarrow \overline{x}$.

In the standard HSA one sets the renormalization scale to be $\mu
_{R}^{2}=Q^{2}$, fixing $\alpha _{\mathrm{s}}(\mu _{R}^{2})$ with
respect to $x$, and simplifying the calculation of the VF \
considerably. In this approach the functions (\ref{eq:1.17}) and
(\ref{eq:1.18}) possess the symmetry features
\begin{equation*}
T_{1(2)}^{q}(x,Q^{2},-\omega ,\eta )=T_{2(1)}^{q}(x,Q^{2},\omega
,\eta ),
\end{equation*}
\begin{equation}
T_{1(2)}^{g}(x,Q^{2},-\omega ,\eta )=-T_{2(1)}^{g}(x,Q^{2},\omega
,\eta ). \label{eq:1.19}
\end{equation}
Using these features, as well as (\ref{eq:1.5}) and the symmetry
properties of $T_{1(2)}^{q(g)}$ under $x\leftrightarrow
\overline{x}$ exchange, one can prove that
\begin{equation}
F_{\eta ^{\prime }g^{\ast }g^{\ast }}^{q(g)}(Q^{2},\omega ,\eta )=F_{\eta
^{\prime }g^{\ast }g^{\ast }}^{q(g)}(Q^{2},-\omega ,\eta ).  \label{eq:1.20}
\end{equation}%
The last equality is a manifestation of the Bose symmetry of the
process under discussion under vector particles--gluons exchange.

Within the RC method the renormalization scale $\mu _{R}^{2}$ \ is chosen
equal, as a rule, to the momentum squared $\ |q^{2}|$ of the virtual partons
in the corresponding Feynman diagrams. For the massless $\eta ^{\prime }$%
-meson on-shell gluon transition the scale $\mu _{R}^{2}$ is exactly equal to

\begin{equation}
\mu _{R}^{2}=Q^{2}x.  \label{eq:1.21}
\end{equation}
Then upon $x\leftrightarrow \overline{x}$ exchange, the argument
of $\alpha _{\mathrm{s}}(\overline{\mu }_{R}^{2})$ in the
functions $T_{2}^{q(g)} $ becomes equal to $\overline{\mu
}_{R}^{2}=Q^{2}\overline{x}$. In the general case, the absolute
value of the square of the four-momenta $q^{2}$ of the virtual
partons depends on the total gluon virtuality $Q^{2}$, the
asymmetry parameter $\omega $ and the scaled mass term $\eta $.
However, to avoid problems related to the appearance of the
parameters $Q^{2}$, $\omega $ and $\eta $ in the argument of
$\alpha _{\mathrm{s}}$, we shall use the renormalization scale
(\ref{eq:1.21}). Such a choice is justified from the physics point
of view as well, because namely the part $\sim Q^{2}x$ of the
renormalization scale leads to the power corrections $\sim
(\Lambda ^{2}/Q^{2})^{n},\,n=1,\,2,...$ to the VF $F_{\eta
^{\prime }g^{\ast }g^{\ast }}(Q^{2},\omega ,\eta )$, which we are
going to compute.

In the present paper we adopt the symmetrized RC method, where $\alpha _{%
\mathrm{s}}(Q^{2}x)$ and $\alpha _{\mathrm{s}}(Q^{2}\overline{x})$ are
replaced by%
\begin{equation*}
\frac{\alpha _{\mathrm{s}}(Q^{2}x)+\alpha _{\mathrm{s}}(Q^{2}\overline{x})}{2%
}.
\end{equation*}%
After this modification all symmetry properties of the
hard-scattering amplitudes and the vertex function remain valid
within the RC method as well. The symmetrized RC method was
successfully employed in the investigation of various exclusive
processes \cite{Ag02,Ag04,AGP}.

\bigskip

\section{Quark component of the vertex function $\ F_{\protect\eta^{\prime}
g^{\ast }g^{\ast }}^{q}(Q^{2},\protect\omega ,\protect\eta)$}

\label{sec:quark}

The expression for the quark component of the vertex function
$F_{\eta ^{\prime }g^{\ast }g^{\ast }}^{q}(Q^{2},\omega ,\eta )$
can be computed only after resolving the problems of the soft
end-point regions $x\rightarrow 0,1$. In fact, having inserted the
explicit expressions of the hard-scattering amplitude and the
quark component of the $\eta ^{\prime }$-meson DA into Eq.\
(\ref{eq:1.15}), one encounters divergences, arising from the
singularities of the coupling constants $\alpha _{\mathrm{s}}(Q^{2}x)$ and $%
\alpha _{\mathrm{s}}(Q^{2}\overline{x})$ in the limits $x\rightarrow 0,1$.
The RC method provides the required prescription to cure these divergences.

To this end, we express the running coupling \\ $\alpha
_{\mathrm{s}}(Q^{2}x)$ in terms of $\alpha _{\mathrm{s}}(Q^{2})$
and, as a result, obtain integrals that can be regularized and
calculated using the approach described in \cite{Ag95,Ag9501}.
Then the quark component of the VF is written as a perturbative
series in $\alpha _{\mathrm{s}}(Q^{2})$ with factorially growing
coefficients. The resummation of such a series is performed by
means of a Borel transformation. Namely, one has to determine the
Borel transform of the corresponding series and in order to get
the resummed expression for the vertex function one has to invert
the Borel transform. The Borel transform of the series with
factorially growing coefficients contains infrared renormalon
poles located at the positive axis of the Borel plane; therefore
the inverse Borel transformation suffers from the pole
divergences. In other words, the Borel technique transforms the
end-point divergences into the IR renormalon pole divergences of
the inverse transformation. Then, the resummed expression can be
extracted by computing the relevant integrals in the sense of the
Cauchy principal value \cite{Ben,Ben01}.

A useful way to bypass these intermediate operations and directly
obtain the Borel resummed expressions is to introduce the
following formula for $\alpha _{\mathrm{s}}(Q^{2}x)$ \cite{Ag01}:

\begin{equation}
\alpha _{\mathrm{s}}(Q^{2}x)=\frac{4\pi }{\beta _{0}}\int_{0}^{\infty
}du\exp (-ut)R(u,t)x^{-u},  \label{eq:3.1}
\end{equation}%
where the function $R(u,t)$ is defined as

\begin{equation*}
R(u,t)=1-\frac{2\beta _{1}}{\beta _{0}^{2}}u(1-\gamma -\ln t-\ln u).
\end{equation*}%
In the above, $\gamma \simeq 0.577216$ is the Euler constant and $t=\ln
(Q^{2}/\Lambda ^{2})$.

Calculations of the quark component of the VF lead to the following result:
\begin{equation}
F_{\eta ^{\prime }g^{\ast }g^{\ast }}^{q}(Q^{2},\omega ,\eta
)=F_{1}^{q}(Q^{2},\omega ,\eta )+F_{2}^{q}(Q^{2},\omega ,\eta ),
\label{eq:3.2}
\end{equation}
where

\begin{equation*}
F_{i}^{q}(Q^{2},\omega ,\eta )=-\frac{16\pi ^{2}C[1+A(Q^{2})]}{Q^{2}\beta
_{0}}K_{i}(\omega ,\eta )
\end{equation*}%
\begin{equation*}
\times \int_{0}^{\infty }due^{-ut}R(u,t)B(2-u,2)\left[ _{2}F_{1}\left(
1,2;4-u;r_{i}\right) \right.
\end{equation*}

\begin{equation*}
\left. +{}_{2}F_{1}\left( 1,2-u;4-u;r_{i}\right) \right] +\frac{80\pi
^{2}CA(Q^{2})}{Q^{2}\beta _{0}}K_{i}(\omega ,\eta )
\end{equation*}

\begin{equation*}
\times \int_{0}^{\infty }due^{-ut}R(u,t)B(3-u,3)\left[ {}_{2}F_{1}\left(
1,3;6-u;r_{i}\right) \right.
\end{equation*}

\begin{equation}
\left. +{}_{2}F_{1}\left( 1,3-u;6-u;r_{i}\right) \right] .  \label{eq:3.3}
\end{equation}%
Here
\begin{equation}
K_{1}(\omega ,\eta )=\frac{\eta }{\omega \lambda (\omega \lambda -\omega
-\eta )},\ r_{1}=\frac{2\eta }{\omega +\eta -\omega \lambda },
\label{eq:3.4}
\end{equation}%
and
\begin{equation}
K_{2}(\omega ,\eta )=\frac{\eta }{\omega \lambda (\omega \lambda -\omega
+\eta )},\ r_{2}=\frac{2\eta }{\eta -\omega +\omega \lambda }.
\label{eq:3.5}
\end{equation}%
Equation (\ref{eq:3.2})  with the functions $F_{1}^{q}$ and
$F_{2}^{q}$ is the Borel resummed expression for the quark
component of the $\eta ^{\prime }g^{\ast }g^{\ast }$ vertex
function and contains power-suppressed corrections coming from the
soft end-point regions \cite{Ag02}.

Since under the replacement $\omega \leftrightarrow -\omega $ the
relations $K_{1}\leftrightarrow K_{2}$ and $r_{1}\leftrightarrow
r_{2}$ hold, we get

\begin{equation*}
F_{\eta ^{\prime }g^{\ast }g^{\ast }}^{q}(Q^{2},-\omega ,\eta )=F_{\eta
^{\prime }g^{\ast }g^{\ast }}^{q}(Q^{2},\omega ,\eta ).
\end{equation*}%
The components $F_{1}^{q}(Q^{2},\omega ,\eta )$ and
$F_{2}^{q}(Q^{2},\omega ,\eta )$ of\\
$F_{\eta ^{\prime}g^{\ast}g^{\ast }}^{q}(Q^{2},\omega ,\eta )$ can
be obtained from each other utilizing the transformation

\begin{equation}
{}_{2}F_{1}(a,b;c;z)=(1-z)^{-a}{}_{2}F_{1}\left( a,c-b;c;\frac{z}{z-1}%
\right) .  \label{eq:3.6}
\end{equation}%
The argument $r_{1}$ of the hypergeometric functions in $F_{1}^{q}(Q^{2},%
\omega ,\eta )$ in the region $\omega \in (-1,0)$ obeys the constraint $%
r_{1}<1$, whereas $r_{2}<1$ in the domain $\omega \in \lbrack
0,1)$. In order to reveal the IR renormalon structure of the Borel
resummed vertex function, as well as to perform numerical
computations, we have to expand the hypergeometric functions over
$r_{1\text{ }}$or $r_{2}$. We choose to work in the region $\omega
\in \lbrack 0,1)$, and we therefore employ the expression
\begin{equation}
F_{\eta ^{\prime }g^{\ast }g^{\ast }}^{q}(Q^{2},\omega ,\eta
)=2F_{2}^{q}(Q^{2},\omega ,\eta ).  \label{eq:3.7}
\end{equation}%
Due to the symmetry of $F_{\eta ^{\prime }g^{\ast }g^{\ast
}}^{q}(Q^{2},\omega ,\eta )$ under $\omega \leftrightarrow -\omega
$, the pole structure of the vertex function and numerical results
in the region $ \omega \in (-1,0)$ are the same.

For the $\eta ^{\prime }$-meson on-shell gluon transition, we get

\begin{equation*}
F_{\eta ^{\prime }gg^{\ast }}^{q}(Q^{2},\omega =\pm 1,\eta )=-\frac{16\pi
^{2}C[1+A(Q^{2})]}{Q^{2}\beta _{0}}\frac{1}{1+\eta }
\end{equation*}%
\begin{equation*}
\times \int_{0}^{\infty }due^{-ut}R(u,t)\left[ B(1,2-u)\right.
\end{equation*}%
\begin{equation*}
\left. +B(2,1-u)\right] +\frac{80\pi ^{2}CA(Q^{2})}{\beta _{0}}\frac{1}{%
1+\eta }
\end{equation*}%
\begin{equation}
\times \int_{0}^{\infty }due^{-ut}R(u,t)\left[ B(3,2-u)+B(2,3-u)\right].
\label{eq:3.8}
\end{equation}
In the case of gluons with equal virtualities, $\omega =0$, the
$F_{\eta ^{\prime }g^{\ast }g^{\ast }}^{q}(Q^{2},\omega =0,\eta )$
can be obtained from (\ref{eq:3.7}) upon the substitutions
$K_{2}\rightarrow K_{0}$ and $r_{2}\rightarrow r_{0}$, where
\begin{equation*}
K_{0}(\eta )=\frac{1}{\eta \sqrt{1+2/\eta }(1+\sqrt{1+2/\eta })},
\end{equation*}
\begin{equation}
\ r_{0}=\frac{2}{1+\sqrt{1+2/\eta }}.  \label{eq:3.9}
\end{equation}

As we have mentioned earlier, the massless $\eta ^{\prime }$-meson
virtual gluon transition vertex function was computed in
\cite{Ag02}. The predictions of this work for $F_{\eta ^{\prime
}g^{\ast}g^{\ast }}^{q}(Q^{2},\omega ,\eta )$ should lead to the
results of \cite{Ag02} in the limit of $\eta \rightarrow 0$. To
regain these results, it is necessary to expand the relevant
functions over $\eta \ll 1$, and only after that take the limit
$\eta \rightarrow 0$. Then, in the general case ($\omega \neq
0,\pm 1$)
\begin{equation*}
K_{2}(\omega ,\eta \rightarrow 0)=\frac{1}{1+\omega },\ \ \ r_{2}=\frac{%
2\omega }{1+\omega }.
\end{equation*}%
Using the last expressions, it is not difficult to check that
(\ref {eq:3.7}) coincides with (4.4) of \cite{Ag02}. By setting
$\eta =0 $, from (\ref{eq:3.8}) one can easily recover the
expression for the massless $\eta ^{\prime }$-meson on-shell gluon
transition VF derived in (4.5) of \cite{Ag02}.

The important question to be clarified here is whether one can use
the results obtained within the RC method in the limit
$Q^{2}\rightarrow \infty $ in order to regain the asymptotic form
of the VF. Indeed, regardless of the method used, in the limit
$Q^{2}\rightarrow \infty $, the VF must reach its asymptotic form,
because power-suppressed corrections vanish in the asymptotic
limit. In the limit $Q^{2}\rightarrow \infty $, the gluon
component of the $\eta ^{\prime }$-meson DA vanishes, $\phi
^{g}(x,Q^{2})\rightarrow 0$, whereas the quark component $\phi
^{q}(x,Q^{2})
$ evolves to its asymptotic form%
\begin{equation*}
\phi ^{q}(x,Q^{2})\rightarrow 6Cx\overline{x}.
\end{equation*}%
Therefore, the results that we obtain here not only describe the
asymptotic limit of the quark component of the VF, but the
asymptotic limit of the full VF itself.

From the whole analysis performed in \cite{Ag02}, it follows that
in the asymptotic limit the substitution
\begin{equation*}
\frac{4\pi }{\beta _{0}}\int_{0}^{\infty }due^{-ut}R(u,t)B(n,m-u)
\end{equation*}%
\begin{equation*}
\times _{2}F_{1}\left( 1,m-u;m+n-u;r\right)
\end{equation*}%
\begin{equation*}
\rightarrow \alpha _{\mathrm{s}}(Q^{2})B(n,m)_{2}F_{1}\left(
1,m;m+n;r\right) ,
\end{equation*}%
has to be applied.

Having used this prescription, we obtain
\begin{equation*}
F_{\eta ^{\prime }g^{\ast }g^{\ast }}^{q}(Q^{2},\omega ,\eta
)\longrightarrow -\frac{8\pi C\alpha _{\mathrm{s}}(Q^{2})}{3Q^{2}}%
K_{i}(\omega ,\eta )
\end{equation*}
\begin{equation}
\times _{2}F_{1}(1,2;4;r_{i}).  \label{eq:3.10}
\end{equation}
Equation (\ref{eq:3.10}) with the quantities labeled $i=1,2$, in
general, should be employed in the relevant regions of the
asymmetry parameter, i.e. in the regions $\omega \in (-1,0)$ and
$\omega \in (0,1)$, respectively. But, because the hypergeometric
function $_{2}F_{1}(1,2,4;z)$ is expressible
in terms of the elementary ones, one can use (\ref{eq:3.10}) with both $%
i=1$ and $i=2$ in the whole region $\omega \in (-1,1)$, excluding the point $%
\omega =0$. At $\omega =1$ ($\omega =-1$) (\ref{eq:3.10}) with
$i=2$ ($ i=1$) can be applied. Our formula for the asymptotic
limit of the quark component of the VF numerically is identical to
(60) of \cite {Ali03} (after setting there $\sqrt{n_{f}}f_{\eta
^{\prime }}\rightarrow C,\
A_{2}(Q^{2})=0$ and evolving the argument of $\alpha _{\mathrm{s}}$ to $%
Q^{2} $).

The IR renormalon structure of the expressions (\ref{eq:3.2}) and (\ref%
{eq:3.8}) does not differ from that of the corresponding massless
$\eta ^{\prime }$-meson gluon transition FFs described in rather
detailed form in \cite{Ag02}, to which we refer interested
readers.

\bigskip

\section{The gluon component of the vertex function}

\label{sec:gluon}

We compute the gluon component of the VF,\\
$F_{\eta ^{\prime }g^{\ast }g^{\ast }}^{g}(Q^{2},\omega ,\eta )$,
employing the formula
\begin{equation*}
F_{\eta ^{\prime }g^{\ast }g^{\ast }}^{g}(Q^{2},\omega ,\eta
)=2\int_{0}^{1}T_{1}^{g}(x,Q^{2},\omega ,\eta )\phi ^{g}(x,Q^{2})dx,
\end{equation*}%
which leads to the following result:

\begin{equation}
F_{\eta ^{\prime }g^{\ast }g^{\ast }}^{g}(Q^{2},\omega ,\eta
)=F_{a}^{g}(Q^{2},\omega ,\eta )+F_{b}^{g}(Q^{2},\omega ,\eta ).
\label{eq:4.1}
\end{equation}%
In (\ref{eq:4.1}) the $a$ and $b$ components are given by the
expressions

\begin{equation*}
F_{a}^{g}(Q^{2},\omega ,\eta )=\frac{4\pi ^{2}CB(Q^{2})}{Q^{2}\beta _{0}n_{f}%
}\frac{\eta }{(\omega \lambda )^{2}(\omega +\eta +\omega \lambda )}
\end{equation*}%
\begin{equation*}
\times \int_{0}^{\infty }due^{-ut}R(u,t)\left\{ (1+\omega )\left[
B(4-u,2)\times {}\right. \right.
\end{equation*}%
\begin{equation*}
\left. _{2}F_{1}(1,2;6-u;\overline{r})+B(4,2-u){}_{2}F_{1}\left( 1,2-u;6-u;\overline{r}%
\right) \right]
\end{equation*}%
\begin{equation*}
-(1-\omega )\left[ B(4,2-u){}_{2}F_{1}\left( 1,4;6-u;\overline{r}\right)
\right.
\end{equation*}%
\begin{equation*}
\left. +B(4-u,2){}_{2}F_{1}\left( 1,4-u;6-u;\overline{r}\right) \right]
\end{equation*}%
\begin{equation*}
-2\omega B(3,3-u){}\left[ _{2}F_{1}\left( 1,3-u;6-u;\overline{r}\right)
\right.
\end{equation*}%
\begin{equation*}
\left. +{}_{2}F_{1}\left( 1,3;6-u;\overline{r}\right) \right]
\end{equation*}%
\begin{equation*}
+2\eta \left[ B(2,3-u){}\left( _{2}F_{1}\left( 1,2;5-u;\overline{r}\right)
\right. \right.
\end{equation*}%
\begin{equation*}
\left. -{}_{2}F_{1}\left( 1,3-u;5-u;\overline{r}\right) \right) +B(3,2-u)
\end{equation*}%
\begin{equation*}
\times \left( _{2}F_{1}\left( 1,2-u;5-u;\overline{r}\right)
-{}_{2}F_{1}\left( 1,3;5-u;\overline{r}\right) \right)
\end{equation*}%
\begin{equation*}
+B(4-u,3)\left( _{2}F_{1}\left( 1,3;7-u;\overline{r}\right) \right.
\end{equation*}%
\begin{equation*}
\left. -{}_{2}F_{1}\left( 1,4-u;7-u;\overline{r}\right) \right) +B(4,3-u)
\end{equation*}%
\begin{equation}
\left. \left. \times \left( _{2}F_{1}\left( 1,3-u;7-u;\overline{r}\right)
-{}_{2}F_{1}\left( 1,4;7-u;\overline{r}\right) \right) \right] \right\} ,
\label{eq:4.2}
\end{equation}%
where
\begin{equation*}
\overline{r}=\frac{2\eta }{\omega +\eta +\omega \lambda },
\end{equation*}%
and

\begin{equation*}
F_{b}^{g}(Q^{2},\omega ,\eta )=\frac{4\pi ^{2}CB(Q^{2})}{Q^{2}\beta _{0}n_{f}%
}\frac{\eta }{(\omega \lambda )^{2}(\eta -\omega +\omega \lambda )}
\end{equation*}%
\begin{equation*}
\times \int_{0}^{\infty }due^{-ut}R(u,t)\left\{ (1+\omega )\left[
B(4-u,2)\times {}\right. \right.
\end{equation*}%
\begin{equation*}
\left. _{2}F_{1}\left( 1,4-u;6-u;r\right) +B(4,2-u){}_{2}F_{1}\left(
1,4;6-u;r\right) \right]
\end{equation*}%
\begin{equation*}
-(1-\omega )\left[ B(4,2-u){}_{2}F_{1}\left( 1,2-u;6-u;r\right) \right.
\end{equation*}%
\begin{equation*}
\left. +B(4-u,2){}_{2}F_{1}\left( 1,2;6-u;r\right) \right]
\end{equation*}%
\begin{equation*}
-2\omega B(3,3-u)\left[ _{2}F_{1}\left( 1,3-u;6-u;r\right) \right.
\end{equation*}%
\begin{equation*}
\left. +{}_{2}F_{1}\left( 1,3;6-u;r\right) \right] +2\eta \left[
B(2,3-u)\times \right.
\end{equation*}%
\begin{equation*}
(_{2}F_{1}\left( 1,3-u;5-u;r\right) -{}_{2}F_{1}\left( 1,2;5-u;r\right) )
\end{equation*}%
\begin{equation*}
+B(3,2-u)\left( _{2}F_{1}\left( 1,3;5-u;r\right) \right.
\end{equation*}%
\begin{equation*}
\left. -{}_{2}F_{1}\left( 1,2-u;5-u;r\right) \right) +B(4-u,3)
\end{equation*}%
\begin{equation*}
\times (_{2}F_{1}\left( 1,4-u;7-u;r\right) -{}_{2}F_{1}\left(
1,3;7-u;r\right) )
\end{equation*}%
\begin{equation*}
+B(4,3-u)\left( {}_{2}F_{1}\left( 1,4;7-u;r\right) \right.
\end{equation*}
\begin{equation}
\left. \left. \left. -{}_{2}F_{1}\left( 1,3-u;7-u;r\right) \right) \right]
\right\} ,  \label{eq:4.3}
\end{equation}
with $r\equiv r_{2}$.

It is worth noting that $F_{a}^{g}(Q^{2},\omega ,\eta )$ has been
obtained from the original result after the transformation
(\ref{eq:3.6}). In the region $\omega \in \lbrack 0,1)$ both $r$
and $\overline{r}<1$, where (\ref{eq:4.2}) and (\ref{eq:4.3}) can
be used for expansion and numerical calculations. We have checked
that,
\begin{equation*}
F_{\eta ^{\prime }g^{\ast }g^{\ast }}^{g}(Q^{2},-\omega ,\eta )=F_{\eta
^{\prime }g^{\ast }g^{\ast }}^{g}(Q^{2},\omega ,\eta ),
\end{equation*}%
which becomes evident if we represent the gluon component of the
VF in the
form%
\begin{equation*}
F_{\eta ^{\prime }g^{\ast }g^{\ast }}^{g}(Q^{2},\omega ,\eta
)=\int_{0}^{1}T_{1}^{g}(x,Q^{2},\omega ,\eta )\phi ^{g}(x,Q^{2})dx
\end{equation*}%
\begin{equation*}
-\int_{0}^{1}T_{2}^{g}(x,Q^{2},\omega ,\eta )\phi ^{g}(x,Q^{2})dx.
\end{equation*}%
But expressions obtained using $T_{2}^{g}(x,Q^{2},\omega ,\eta )$
are as lengthy as the ones presented in (\ref{eq:4.2}) and
(\ref{eq:4.3}); therefore, we refrain from writing them down here.

In the case of the $\eta ^{\prime }$-meson on-shell gluon
transition, the VF can be found after changing the factors and
arguments of the hypergeometric functions in (\ref{eq:4.2}) and
(\ref{eq:4.3}), i.e. \newline
in $F_{a}^{g}(Q^{2},\omega ,\eta )$%
\begin{equation*}
\frac{\eta }{(\omega \lambda )^{2}(\omega +\eta +\omega \lambda )}%
\rightarrow \frac{\eta }{2(1+\eta )^{3}},\ \ \overline{r}\rightarrow \frac{%
\eta }{1+\eta },
\end{equation*}%
in $F_{b}^{g}(Q^{2},\omega ,\eta )$%
\begin{equation*}
\frac{\eta }{(\omega \lambda )^{2}(\eta -\omega +\omega \lambda )}%
\rightarrow \frac{1}{2(1+\eta )^{2}},\ \ r\rightarrow 1.
\end{equation*}%
The gluon component of the VF is identically equal to zero for equal gluon
virtualities ($\omega =0$). This is the important qualitative modification
induced by the $\eta ^{\prime }$-meson mass term kept in the hard-scattering
amplitudes. Let us emphasize that $F_{\eta ^{\prime }g^{\ast }g^{\ast
}}^{g}(Q^{2},\omega =0,\eta )\equiv 0$ within both the standard HSA and the
RC method.

In the limit $\eta \rightarrow 0$ our results reproduce the
expression for the gluon component of the massless $\eta ^{\prime
}$-meson virtual gluon transition VF from (4.22) of \cite{Ag02}.
In fact, acting as in the case of the quark component of VF, we
can show that the factor in $F_{a}^{g}(Q^{2},\omega ,\eta )$
vanishes,
\begin{equation*}
\frac{\eta }{(\omega \lambda )^{2}(\omega +\eta +\omega \lambda )}%
\rightarrow 0,
\end{equation*}%
and for $F_{b}^{g}(Q^{2},\omega ,\eta )$ get:
\begin{equation*}
\frac{\eta }{(\omega \lambda )^{2}(\eta -\omega +\omega \lambda )}%
\rightarrow \frac{1}{\omega (1+\omega )},\ \ r\rightarrow \frac{2\omega }{%
1+\omega }.
\end{equation*}%
Then it is easy to demonstrate that, the function
$F_{b}^{g}(Q^{2},\omega ,\eta )$ in the limit $\eta \rightarrow 0$ leads to $%
F_{\eta ^{\prime }g^{\ast }g^{\ast }}^{g}(Q^{2},\omega )$.

The infrared renormalon structure of the terms in (\ref{eq:4.2})
and (\ref{eq:4.3}) $\sim (1+\omega ),\ (1-\omega ),\ 2\omega $ are
the same as in the case of the massless $\eta ^{\prime }$-meson
virtual gluon transition VF. The terms $\sim 2\eta $ are new;
nevertheless, their IR renormalon structures can be clarified
using the procedures described in \cite{Ag02}.

%%%%%%%%%%%%%%%%%%%%%%%%%%%%%%%%%%%%%%%%%%%%%%%%%%%%%%%%%%%%%%%%%%%%%%%%%%%%%%%%
\begin{figure}[t]
%%Fif 3a & Fig 3b
%
\centering\epsfig{file=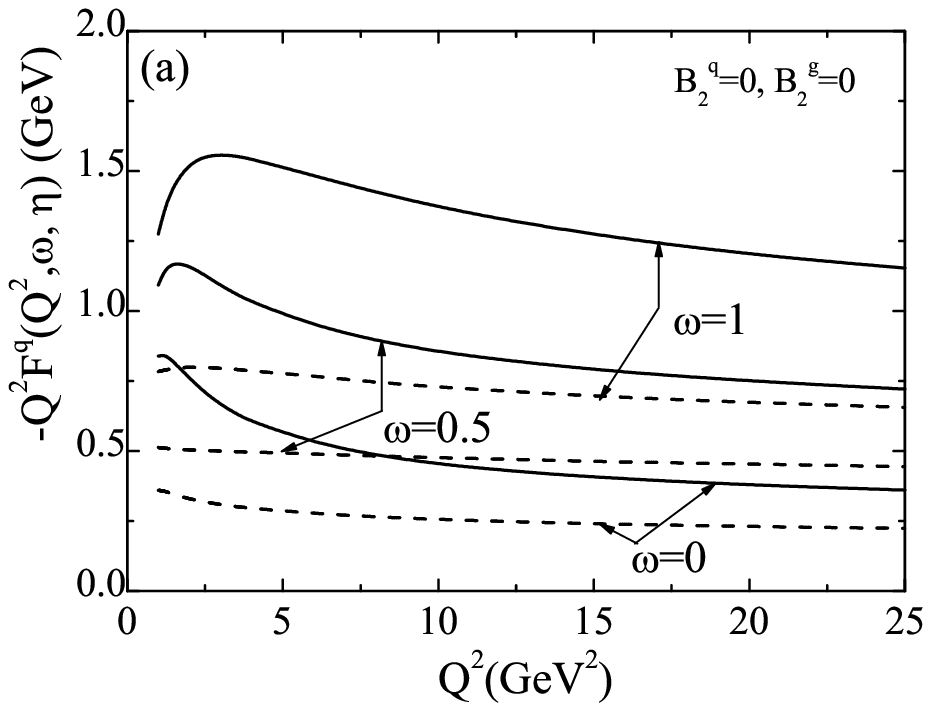,height=6cm,width=8.cm,clip=}
\vspace{0.0cm}
\centering\epsfig{file=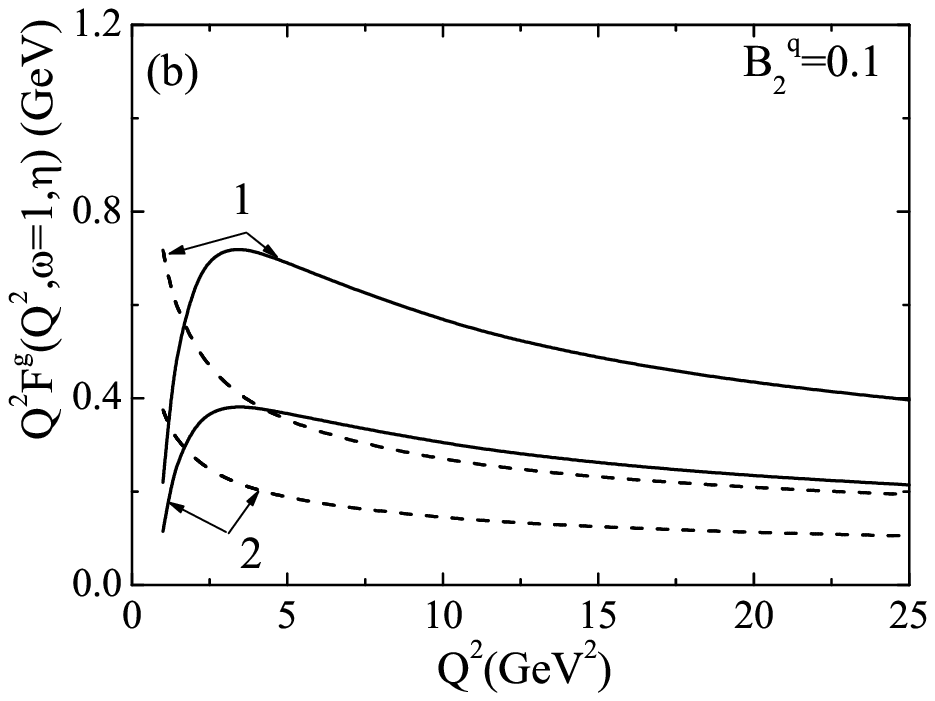,height=6cm,width=8.cm,clip=}
\vspace{-0.6cm} \caption{ The quark ({\bf a}) and gluon ({\bf b})
components of the
scaled VF $Q^{2}F_{\protect\eta ^{\prime }g^{\ast }g^{\ast }}(Q^{2}, \protect%
\omega, \protect\eta )$ as functions of $Q^{2}$. The {\it solid
curves} are obtained using the RC method, whereas the {\it dashed
lines} are calculated within the standard HSA. In {\bf b} the
correspondence between the curves and input parameter $B_{2}^{g}$
is $B_{2}^{g}=8$ for the curves $1$ and $B_{2}^{g}=4$ for the
curves $2$} \label{fig:kfig3}
\end{figure}

\bigskip

\section{Numerical analysis}

\label{sec:num}

%%%%%%%%%%%%%%%%%%%%%%%%%%%%%%%%%%%%%%%%%%%%%%%%%%%%%%%%%%%%%%%%%%%%%
\begin{figure}[t]
%%Fif 4a & Fig 4b
%
\centering\epsfig{file=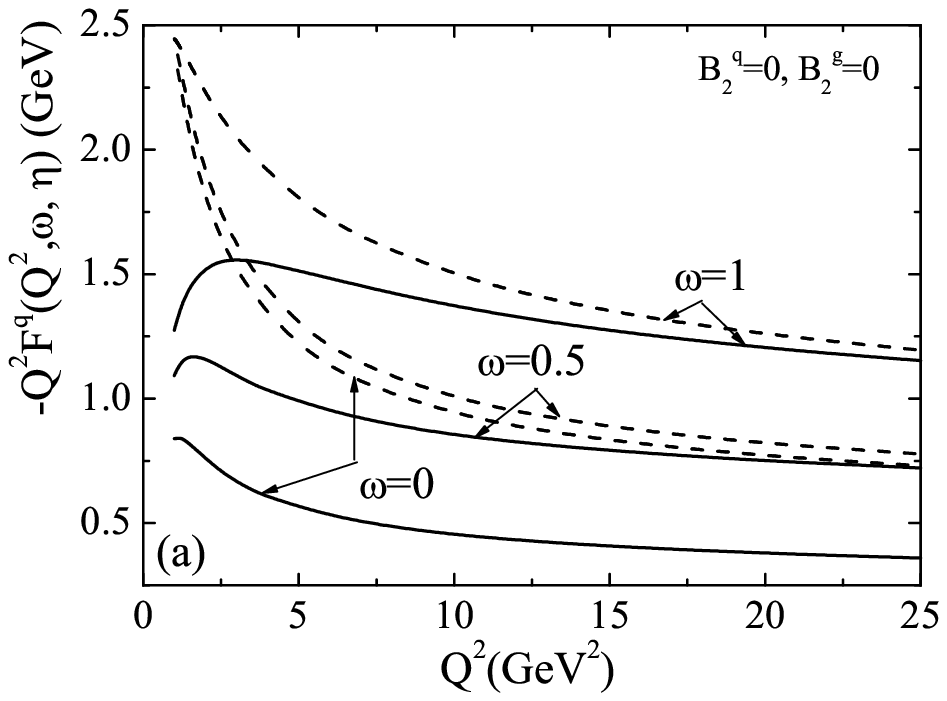,height=6cm,width=8.0cm,clip=} \hspace{0.cm%
} \centering\epsfig{file=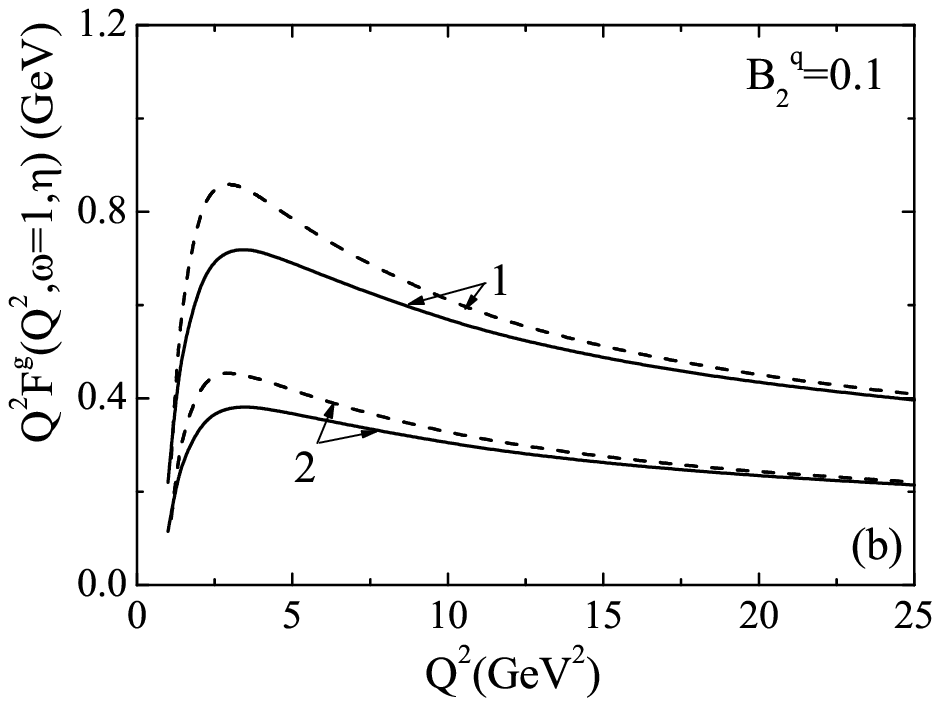,height=6cm,width=8.0cm,clip=} \vspace{%
-0.6cm} \caption{The quark ({\bf a}) and gluon ({\bf b})
components of the VF versus $Q^{2}$. All curves are obtained in
the context of the RC method. The {\it solid lines} are calculated
by taking into account the $\protect\eta^{\prime}$-meson mass
effects: in computations of the {\it dashed lines} the
$\protect\eta^{\prime}$-meson mass term is neglected. In {\bf b}
the correspondence between the curves and parameter $B_{2}^{g}$ is
the same as in Fig.\ \protect\ref{fig:kfig3}} \label{fig:kfig4}
\end{figure}
%%%%%%%%%%%%%%%%%%%%%%%%%%%%%%%%%%%%%%%%%%%%%%%%%%%%%%%%%%%%%%%%%%%%%

%%%%%%%%%%%%%%%%%%%%%%%%%%%%%%%%%%%%%%%%%%%%%%%%%%%%%%%%%%%%%%%%%%%%%
\begin{figure}[t]
%%Fif 5a & Fig 5b
%
\centering\epsfig{file=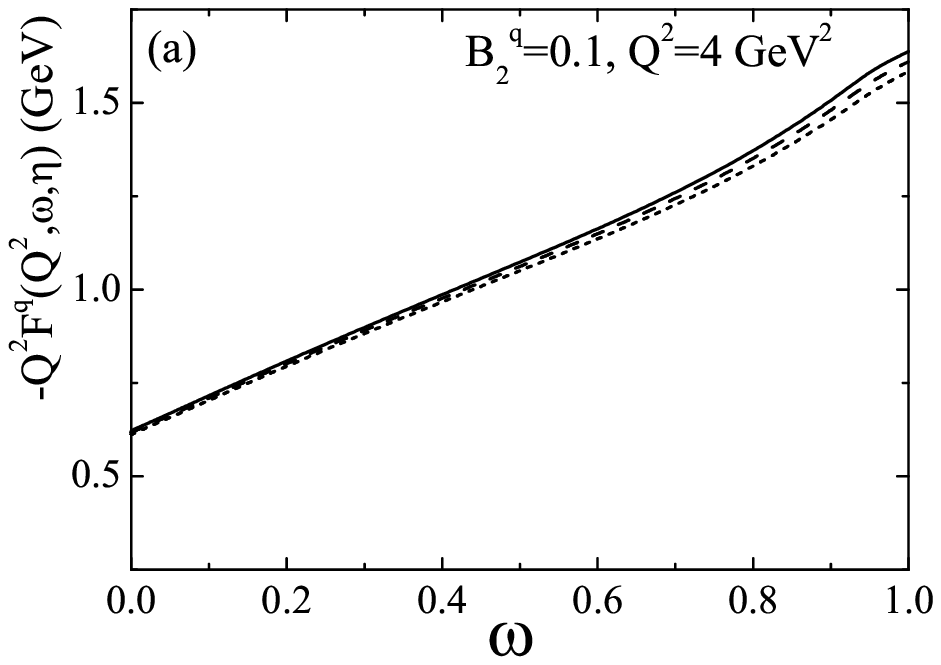,height=6cm,width=8.0cm,clip=} \hspace{0.cm%
} \centering\epsfig{file=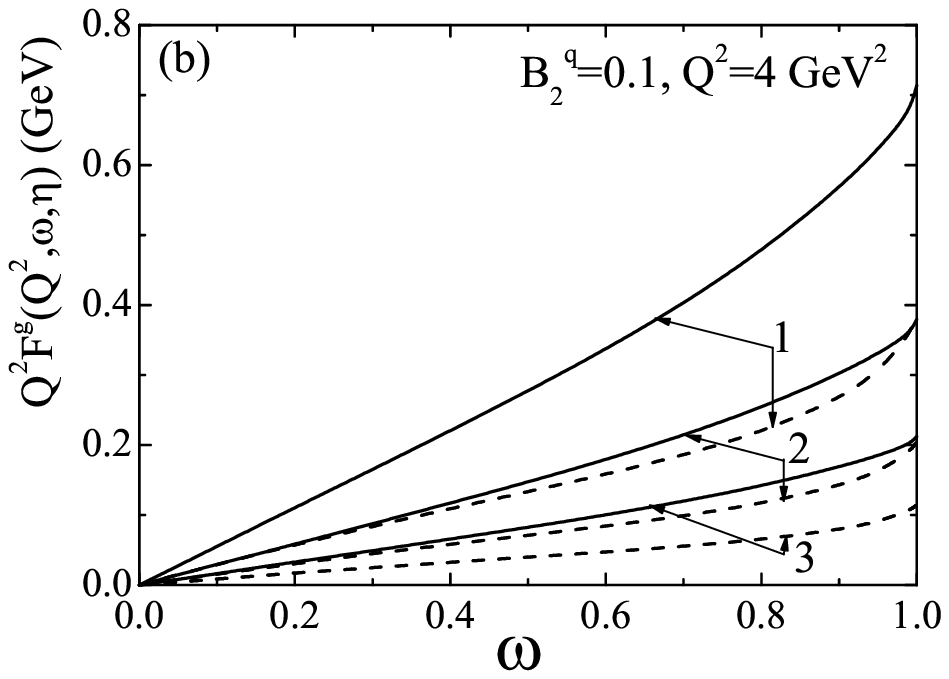,height=6cm,width=8.0cm,clip=}
\vspace{ -0.6cm} \caption{The quark ({\bf a}) and gluon ({\bf b})
components of the VF at fixed $B_2^q$ and $Q^2$ versus
$\protect\omega$. In {\bf a} all curves are obtained in the
context of the RC method. The correspondence between them and the
parameter $B_{2}^{g}$ is: $B_{2}^{g}=0$ for the {\it solid curve};
$B_{2}^{g}=4$ for the {\it dashed curve}; $B_{2}^{g}=8$ for the
{\it short-dashed curve}. In {\bf b} the {\it solid curves} are
obtained within the RC method. For computation of the {\it broken
lines} the standard HSA is used. The correspondence between the
curves and the parameter $B_{2}^{g}$ is $B_{2}^{g}=8$ for the
curves $1$; $B_{2}^{g}=4$ for the curves $2$ and $B_{2}^{g}=2$ for
the lines $3$} \label{fig:kfig5}
\end{figure}
%%%%%%%%%%%%%%%%%%%%%%%%%%%%%%%%%%%%%%%%%%%%%%%%%%%%%%%%%%%%%%%%%%%%%

In order to start numerical computations, we need to fix the values of some
constants and parameters. In our calculations the $\eta ^{\prime }$-meson
mass is set equal to $m_{\eta ^{\prime }}=0.958\ \mathrm{GeV}$. The value of
the QCD scale parameter for $n_{f}=4$ is $\Lambda =0.3\ \mathrm{GeV}$.

To proceed with the computation of the $\eta ^{\prime }$-meson
gluon vertex function and explore the role played by the $\eta
^{\prime }$-meson gluon content and its mass in this process, we
have to define also the allowed values of the input parameters
$B_{2}^{q}$ and $B_{2}^{g}$ at the normalization scale $\mu
_{0}^{2}=1\;\mathrm{GeV}^{2}$. In the present paper we use the
$\eta^{\prime}$-meson asymptotic DA or select values of the
parameters $B_{2}^{q}$ and $B_{2}^{g}$ that obey the constraints
\begin{equation}
B_{2}^{q}=0.1,\;B_{2}^{g}\in \lbrack -2,14].  \label{eq:5.1}
\end{equation}

The quark component of the $\eta ^{\prime }$-meson virtual gluon
transition VF for different values of the asymmetry parameter is
shown in Fig.\ \ref{fig:kfig3}a. The chosen values of the input
parameters correspond to the $\eta ^{\prime }$-meson asymptotic
DA. Since, in the case of the asymptotic DA, the gluon component
of the VF vanishes, in this figure we, actually, have curves for
the full VF. In the same figure predictions obtained within the
standard HSA are also depicted. One sees, that in the domain $1\
\mathrm{\leq Q^{2}\leq 25\ {GeV}^{2}}$ the standard pQCD results
get enhanced by approximately a factor of two due to power
corrections. A similar conclusion is valid also for the gluon
component of the VF for $\mathrm{Q^{2}\geq 4\ {GeV}^{2}}$ ( Fig.\
\ref{fig:kfig3}b) as well.

%%%%%%%%%%%%%%%%%%%%%%%%%%%%%%%%%%%%%%%%%%%%%%%%%%%%%%%%%%%%%%%%%%%%%
%                            F I G U R E  6                         %
%                          \label{fig:kfig6}                        %
%%%%%%%%%%%%%%%%%%%%%%%%%%%%%%%%%%%%%%%%%%%%%%%%%%%%%%%%%%%%%%%%%%%%%
\begin{figure}[t]
%%Fif 6a & Fig 6b
%
\centering\epsfig{file=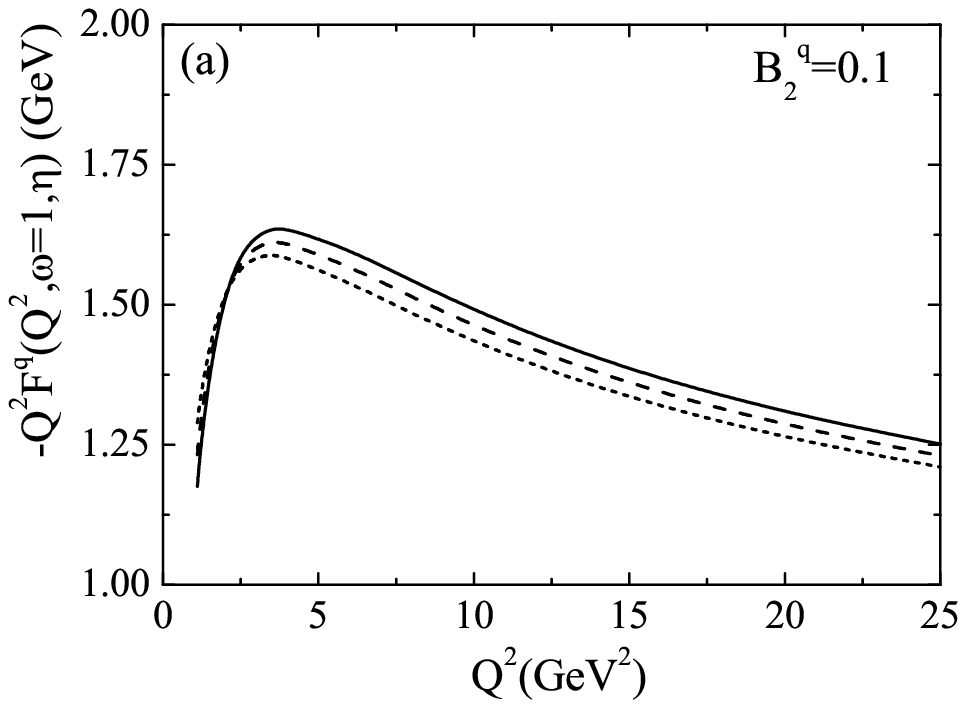,height=6cm,width=8.0cm,clip=}
\hspace{0.cm}
\centering\epsfig{file=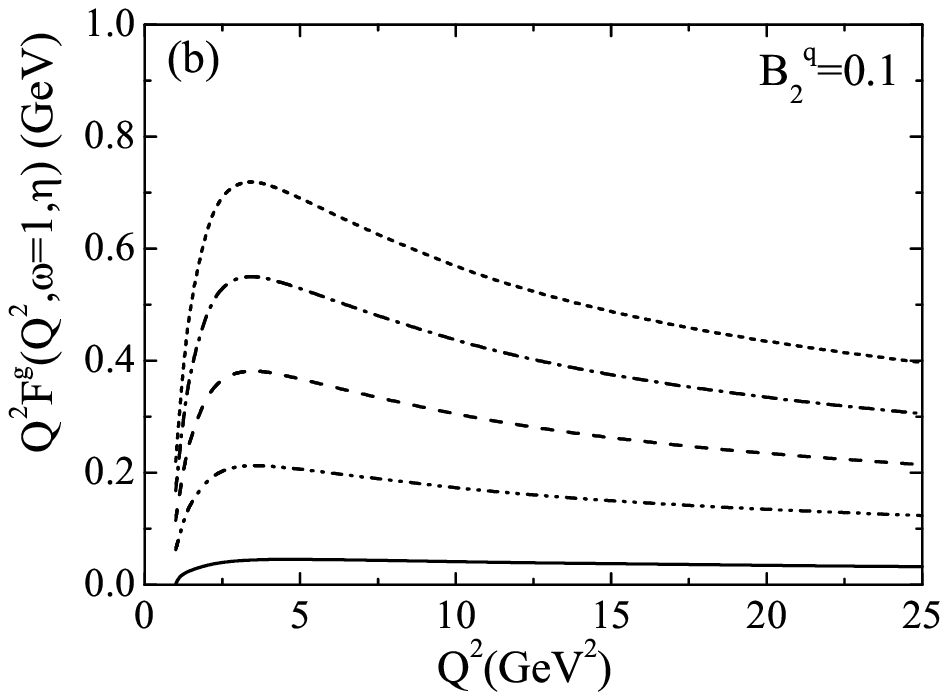,height=6cm,width=8.0cm,clip=}
\vspace{ -0.6cm} \caption{The quark ({\bf a}) and gluon ({\bf b})
components of the VF as functions of $Q^2$ at $\protect\omega= \pm
1$. All curves are computed using the RC method. The
correspondence between plotted curves and parameter $B_2^g$ is
$B_2^g=0$ for the {\it solid curves}; $B_2^g=2$ for the {\it
dot-dot-dashed curves}; $B_2^g=4$ for the {\it dashed lines};
$B_2^g=6$ for the {\it dot-dashed lines} and $B_2^g=8$ for the
{\it short-dashed curves} } \label{fig:kfig6}
\end{figure}
%%%%%%%%%%%%%%%%%%%%%%%%%%%%%%%%%%%%%%%%%%%%%%%%%%%%%%%%%%%%%%%%%%%%%

Even from these first results it is evident that soft end-point
corrections lead, approximately, to the same enhancement of the
standard predictions, as in the case of the massless $\eta
^{\prime }$-meson virtual gluon transition \cite{Ag02}. Therefore,
it is interesting to find modifications in the behavior of the VF
induced by the $\eta ^{\prime }$-meson mass effects.

%%%%%%%%%%%%%%%%%%%%%%%%%%%%%%%%%%%%%%%%%%%%%%%%%%%%%%%%%%%%%%%%%%%%%
%                            F I G U R E  7                         %
%                          \label{fig:kfig7}                        %
%%%%%%%%%%%%%%%%%%%%%%%%%%%%%%%%%%%%%%%%%%%%%%%%%%%%%%%%%%%%%%%%%%%%%
\begin{figure}[t]
%%Fif 7a & Fig 7b
%
\centering\epsfig{file=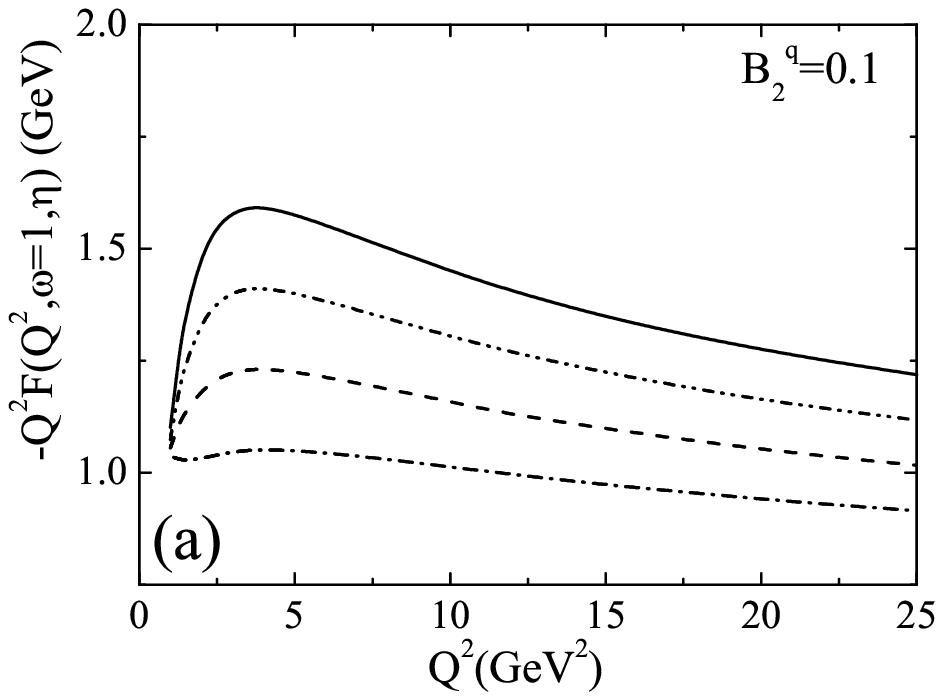,height=6cm,width=8.0cm,clip=}
\hspace{0.cm}
\centering\epsfig{file=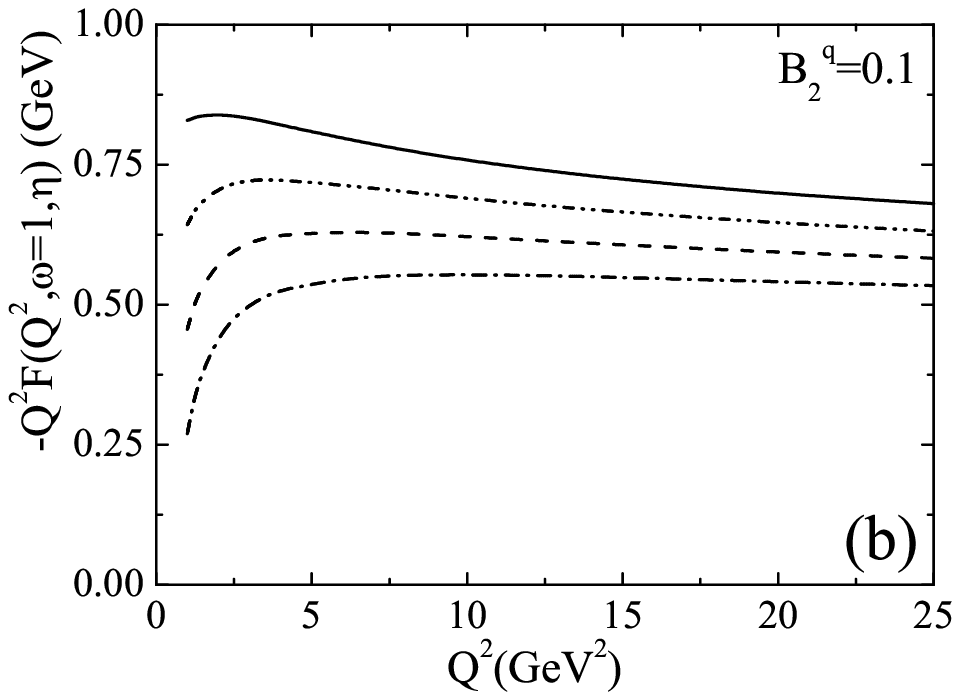,height=6cm,width=8.0cm,clip=}
\vspace{ -0.6cm} \caption{The full VF at $\protect\omega=\pm 1$
computed using the RC method ({\bf a}) and the standard HSA ({\bf
b}). The correspondence between the depicted lines and the
parameter $B_{2}^{g}$ is the same as in Fig.\
\protect\ref{fig:kfig6} (the {\it short-dashed lines} are not
shown)} \label{fig:kfig7}
\end{figure}
%%%%%%%%%%%%%%%%%%%%%%%%%%%%%%%%%%%%%%%%%%%%%%%%%%%%%%%%%%%%%%%%%%%%%

In Fig.\ \ref{fig:kfig4}, the VFs computed within the RC method by
taking into account and neglecting the $\eta ^{\prime }$-meson
mass term, are demonstrated. The differences in behavior of the
quark component (Fig.\ 4$a$) are considerable. Indeed, mass
effects suppress the absolute value of the quark component and, at
the same time, change its shape in the pQCD applicable region of
$Q^{2}$. The gluon component, as a function of the total gluon
virtuality $Q^{2}$, is not affected dramatically by the $\eta
^{\prime }$-meson mass effects (Fig.\ 4b).

But the $\eta ^{\prime }$-meson mass term changes drastically the
behavior of the gluon component of the VF as a function of the
asymmetry parameter. As a function of $\omega $, the gluon
component is plotted in Fig.\ \ref {fig:kfig5}b. It turns out that
$F_{\eta ^{\prime }g^{\ast }g^{\ast }}^{g}(Q^{2},\omega ,\eta )$
vanishes at $\omega =0$ in the framework of both the standard HSA
(the dashed lines) and the RC method (the solid lines). For small
$\omega $ the end-point effects are also mild.
Therefore, it is legitimate to conclude that in the region $|\omega |<0.2$, $%
F_{\eta ^{\prime }g^{\ast }g^{\ast }}^{g}(Q^{2},\omega ,\eta )$
feels neither end-point nor mass effects. The dependence of the
quark component $ F_{\eta ^{\prime }g^{\ast }g^{\ast
}}^{q}(Q^{2},\omega ,\eta )$ on $\omega $ is plotted in panel (a)
of the same figure: as a function of $\omega $ it demonstrates
firm stability against variations of $B_{2}^{g}$.

We have analyzed the impact of the various DAs of the $\eta ^{\prime }$%
-meson on the VF. The quark component of the VF is stable for
different values of $B_{2}^{g}\in \lbrack 0,8]$ (Fig.\
\ref{fig:kfig6}, panel (a)). In contrast, the gluon component of
the VF demonstrates rapid growth with $B_{2}^{g}$ (Fig.\
\ref{fig:kfig6}b). As a result, due to different signs of the
quark and gluon components of the VF, the total vertex function
$F_{\eta ^{\prime }gg^{\ast }}(Q^{2},\omega =\pm 1,\eta )$ for
$B_{2}^{g}\neq 0$ runs below the asymptotic one (Fig.\ 7a). For
comparison the predictions derived in the standard HSA are also
shown (Fig.\ 7b). The quantitative difference between the
corresponding curves is clear.

%%%%%%%%%%%%%%%%%%%%%%%%%%%%%%%%%%%%%%%%%%%%%%%%%%%%%%%%%%%%%%%%%%%%%
%                            F I G U R E  8                        %
%                          \label{fig:kfig8}                        %
%%%%%%%%%%%%%%%%%%%%%%%%%%%%%%%%%%%%%%%%%%%%%%%%%%%%%%%%%%%%%%%%%%%%%
\begin{figure}[tbp]
%%Fig 8
\epsfxsize=8 cm \epsfysize=6 cm \centerline{\epsffile{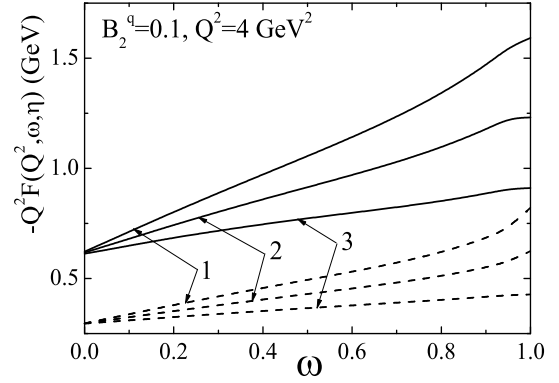}}
\vspace{-0.6cm} \caption{The full VF obtained by employing the RC
method (the {\it solid lines}) and the standard HSA (the {\it
dashed lines}) as a function of $\protect\omega$. For the lines
$1$ the parameter is $B_2^g=0$; for the lines $2$ -- $B_2^g=4$,
and for the lines $3$ -- $B_2^g=8$ } \label{fig:kfig8}
\end{figure}
%%%%%%%%%%%%%%%%%%%%%%%%%%%%%%%%%%%%%%%%%%%%%%%%%%%%%%%%%%%%%%%%%%%%%

The dependence of the full VF on the asymmetry parameter $\omega $ is
depicted in Fig.\ \ref{fig:kfig8}. In the calculations the $\eta ^{\prime }$%
-meson DAs with various values of $B_{2}^{g}$ are employed. As it has been
noted above, the gluon component of the VF is identically equal to zero at $%
\omega =0$, and the quark component, as a function of $\omega $,
demonstrates stability against variations of $B_{2}^{g}$.
Therefore, it is easy to understand the features of the full VF as
a function of $\omega $. Really, in the region $|\omega |<0.3$ the
difference between the VFs
corresponding to different $B_{2}^{g}$ is small; it becomes essential for $%
|\omega |>0.8$. But, owing to the power corrections, mainly to the
quark component, even for $|\omega |<0.3$ the Borel resummed full
VF significantly exceeds the standard pQCD result.

For phenomenological applications it is useful to parameterize the
VF using some simple expressions. We start from the expressions of
the VF obtained in the framework of the standard HSA. For the sake
of simplicity, let us consider the $\eta ^{\prime }$-meson
asymptotic DA. In this case, the standard pQCD prediction for the
VF is given by (\ref{eq:3.10}). Now, we want to approximate the RC
prediction in the form

\begin{equation}
F_{\eta ^{\prime }g^{\ast }g^{\ast }}^{q}(Q^{2},\omega ,\eta )=-\frac{8\pi
C\alpha _{\mathrm{s}}(Q^{2})}{3Q^{2}}K_{2}(\omega ,\eta )  \label{eq:5.2}
\end{equation}%
\begin{equation*}
\times _{2}F_{1}(1,2;4;r_{2})\left( a+\frac{b}{Q^{2}}+\frac{c}{Q^{4}}%
+\ldots \right).
\end{equation*}%
From the general formula (\ref{eq:5.2}), for the $\eta ^{\prime
}$-meson on-shell gluon transition we get

\begin{equation}
F_{\eta ^{\prime }gg^{\ast }}^{q}(Q^{2},\omega =\pm 1,\eta )=-\frac{4\pi
C\alpha _{\mathrm{s}}(Q^{2})}{Q^{2}}  \label{eq:5.3}
\end{equation}
\begin{equation*}
\times \frac{1}{1+\eta }\left(
a+\frac{b}{Q^{2}}+\frac{c}{Q^{4}}+\ldots \right) .
\end{equation*}%
The fitting procedure gives the following values of the
parameters:
\begin{equation*}
a\simeq 1.7837,\ \ b\simeq 0.8228,\ \ c\simeq -1.022.
\end{equation*}

%%%%%%%%%%%%%%%%%%%%%%%%%%%%%%%%%%%%%%%%%%%%%%%%%%%%%%%%%%%%%%%%%%%%%
%                            F I G U R E  9                        %
%                          \label{fig:kfig8}                        %
%%%%%%%%%%%%%%%%%%%%%%%%%%%%%%%%%%%%%%%%%%%%%%%%%%%%%%%%%%%%%%%%%%%%%
\begin{figure}[tbp]
%%Fig 9
\epsfxsize=8 cm \epsfysize=6 cm \centerline{\epsffile{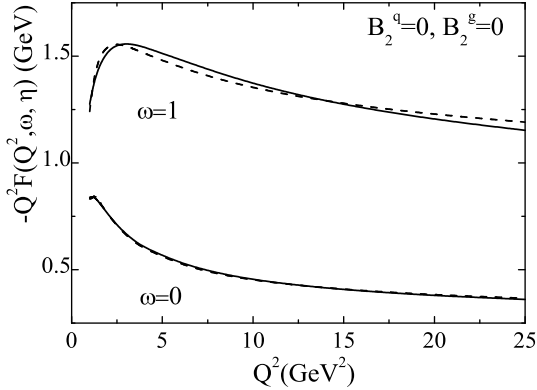}}
\vspace{-0.6cm} \caption{The approximations to the full VF. For
the {\it upper curves} $\protect\omega =1$, for the {\it lower
ones} $\protect\omega =0$. The {\it solid lines} are RC
predictions, the {\it dashed ones} the corresponding
approximations} \label{fig:kfig9}
\end{figure}
%%%%%%%%%%%%%%%%%%%%%%%%%%%%%%%%%%%%%%%%%%%%%%%%%%%%%%%%%%%%%%%%%%%%%

In the case of the gluons with equal virtualities, we find

\begin{equation}
F_{\eta ^{\prime }g^{\ast }g^{\ast }}^{q}(Q^{2},\omega =0,\eta )=-\frac{8\pi
C\alpha _{\mathrm{s}}(Q^{2})}{3Q^{2}}K_{0}(\eta )  \label{eq:5.4}
\end{equation}%
\begin{equation*}
\times _{2}F_{1}(1,2;4;r_{0})\left( a+\frac{b}{Q^{2}}+\frac{c}{Q^{4}}%
+\ldots \right) ,
\end{equation*}%
with

\begin{equation*}
a\simeq 1.5485,\ \ b\simeq 2.3361,\ \ c\simeq -1.627.
\end{equation*}%
The corresponding results are shown in Fig.\ \ref{fig:kfig9}. As
is seen, (\ref{eq:5.4}) leads to an almost perfect approximation
of the original result, whereas the expression (\ref{eq:5.3})
describes the exact prediction, demonstrating, nevertheless, some
deviations.

It is known, that the RC method produces higher-twist ambiguities.
For the massless $\eta^{\prime}$-meson gluon transition VF they
were estimated in \cite{Ag02} and found to lie within $\pm 15 \%$
of the original results. Because such modifications cannot change
our principal conclusions, we do not concentrate on these
questions here.

\section{Concluding remarks}

\label{sec:conc}

In this paper we have evaluated soft end-point (power-suppressed)
corrections to the space-like $\eta^{\prime}$-meson virtual gluon
transition VF by including the $\eta^{\prime}$-meson mass effects.
To this end, we have employed the standard HSA and RC method in
conjunction with the IR renormalon calculus. In the calculations,
both the quark and the gluon components of the
$\eta^{\prime}$-meson DA have been taken into account. We have
modelled the DAs by retaining in the general expressions
(\ref{eq:1.6}) and (\ref{eq:1.7}) only the first non-asymptotic
terms.

We have extended the results obtained in \cite{Ag02} for the
massless $ \eta^{\prime}$-meson virtual gluon transition VF. It
has been shown that effects generated by the $\eta^{\prime}$-meson
mass term considerably change the predictions obtained in
\cite{Ag02}: they suppress the absolute values of the quark and
gluon components of the VF, and modify their behavior as functions
of the asymmetry parameter $\omega$. This modification, in the
case of the gluon component, has not only a quantitative, but also
a qualitative character: thus, at $\omega=0$ the gluon component
of the VF vanishes identically. As a result, mass effects change
the dependence of the full VF on the total gluon virtuality $Q^2$
and asymmetry parameter $\omega$.

The numerical analysis presented shows that power corrections
considerably enhance the standard pQCD predictions for the VF in
the explored region\\
$1\;\mathrm{{GeV}^{2}\leq Q^{2}\leq 25\;{GeV}^{2}}$, though other
sources, may also give rise to power corrections. As an important
consistency check, we have proven that the results obtained within
the RC method in the asymptotic limit $Q^{2}\rightarrow \infty $
reproduce the standard pQCD predictions for the vertex function.

\end{document}